\documentstyle[12pt,aasms4]{article}
\slugcomment{submitted to ApJ}

\def\lsim{\lower.5ex\hbox{$\; \buildrel < \over \sim \;$}}
\def\gsim{\lower.5ex\hbox{$\; \buildrel > \over \sim \;$}} 
\def\lax    {\ifmmode{_<\atop^{\sim}}\else{${_<\atop^{\sim}}$}\fi}
\def\gax    {\ifmmode{_>\atop^{\sim}}\else{${_>\atop^{\sim}}$}\fi}
\def\etal{{\it et al.\/} }
\def\gtorder{\mathrel{\raise.3ex\hbox{$>$}\mkern-14mu
	     \lower0.6ex\hbox{$\sim$}}}
\def\ltorder{\mathrel{\raise.3ex\hbox{$<$}\mkern-14mu
	     \lower0.6ex\hbox{$\sim$}}}

\begin{document}

\title{Temporal and Spectral Properties of Comptonized 
Radiation and Its Applications}

\author{Demosthenes Kazanas, Xin-Min Hua\altaffilmark{1} and 
        Lev Titarchuk\altaffilmark{2}} 
\affil{LHEA, NASA/GSFC Code 661, Greenbelt, MD 20771}

\altaffiltext{1}{NRC/NAS Senior Research Associate}
\altaffiltext{2}{CSI, George Mason University}

%\vskip 0.5 truecm
%\font\rom=cmr10
%\centerline{\rom In press, Astrophys. J. (Letters)}

\begin{abstract}

We have found  relations between the temporal and spectral properties 
of radiation Comptonized in an extended atmosphere associated
with compact accreting sources. We demonstrate that 
the fluctuation power spectrum density (PSD) imposes constraints 
on the atmosphere scale and profile. Furthermore, we indicate that 
the slope and low frequency break of the PSD are related to the
Thomson depth $\tau_0$ of the atmosphere and the radius of its physical 
size respectively. Since the energy spectrum of the escaping radiation
depends also on $\tau_0$ (and the electron temperature $kT_e$),
the relation between spectral and temporal properties follows.
This relation allows for the first time an estimate of the accreting 
matter Thomson depth $\tau_0$ independent of arguments involving 
Comptonization. We present figures for the light curves and PSD 
of different energy bands, the photon energy spectra and the phase 
lags as functions of the variability frequency. 
The temporal properties of the high (soft) and low (hard) state of 
black hole sources are discussed in this context.

\end{abstract}

\keywords{accretion--- black hole physics--- radiation mechanisms: 
Compton and inverse Compton--- stars: neutron--- X-rays: QPO}

\section{Introduction} 

Much of the information about the conditions in the vicinity of compact
accreting objects, the region where most of the accretion kinetic energy
is themalized and converted into radiation, has so far come from
observations and modeling of their energy spectra. Specifically, the
observations of power-law high energy ($E \gsim 2 - 50 $ keV) spectra has
established the process of unsaturated Comptonization of soft photons by
hot electrons as the main process by which these spectra are produced (see
e.g. Sunyaev \& Titarchuk 1980). However, the spectral information alone
does not suffice to determine the physical parameters of the system, since
it can only determine the combination of the Thomson depth, $\tau_0$, 
and the electron temperature, $T_e$, which is relevant to 
the Comptonization of soft photons by the hot electrons, 
namely the Comptonization parameter $y$ defined as  
$y\simeq (4kT_e/m_ec^2)\tau_0^2$, for $\tau_0 \gg 1$.
In this respect, one should bear in mind that for
quasi-spherical, free-fall accretion, the above parameters, 
{\it i.e.} $\tau_0, ~y$, are independent of the mass of the
accreting object,  provided that the accretion proceeds at the
same fraction of its Eddington value. Thus, in order to completely 
determine the physical parameters of  these systems, 
additional, independent information is needed to set the scale of the radius
and the density of the emitting region, information which can only come from
the dynamics of accretion. These arguments have motivated the study of  time 
variability of accreting sources and in particular of galactic black hole 
candidates; it was thought that the absence of a solid
boundary in these bright sources would allow one to probe the dynamics of
accretion onto the black hole. 

However, the study of time variability of black hole candidate sources, in
particular the archetypal galactic source Cyg X--1, has yielded a 
number of results which were rather unexpected when viewed within the 
standard framework of the viscous accretion disk model for
the dynamics of matter accretion  and emission of radiation from these sources: 
(a) The fluctuation power spectral densities (hereafter PSD) are generally 
power laws of indices $s \sim 1 - 1.5$ in the variability frequency 
$\omega$, {\it i.e.,}
$\vert F(\omega)\vert ^2 \propto \omega^{-s} ~{\rm for}~ \omega >
\omega_c$ and $ \vert F(\omega) \vert ^2 \propto \omega^0 ~{\rm for}~
\omega < \omega_c$ (see e.g. van der Klis 1995). 
The values of the slope, $s$, of the PSD are thus significantly 
flatter than these expected from exponential shots associated with 
the turbulent dynamics of the accreting gas near its last stable orbit
($s \ge 2$). Furthermore, the turnover frequency, $\omega_c$, is 
generally several orders of magnitude smaller 
than those associated with the dynamics responsible for the emission of
high energy radiation.  For Cyg X--1, the archetypal galactic black hole
candidate, $s \simeq 1$ (flicker noise), while $\omega_c \simeq 0.05$ Hz,
a far cry from the kHz frequencies expected on the basis of dynamical
considerations. (b) In addition to these broad band spectral features, the
PSD of black hole candidates also exhibit QPO's. The observation of these
features, thought to result from the ``beating"  between the Keplerian
frequency of an accretion disk and the rotation of an underlying 
neutron star, poses a problem for systems thought to contain black holes. 
(c) The spectral -- temporal studies of Miyamoto et al. (1988) of the
time lags between hard and soft photons as a function of the
variability frequency $\omega$, has indicated that these lags, at least   
in Cyg X--1, decrease with increasing variability frequency $\omega$, 
thus ruling out 
Comptonization by an electron cloud of uniform density and temperature. 

Motivated by the discrepancy of the above systematics of time 
variability in black hole candidates from those expected  on the 
basis of the ``standard" dynamical models of viscous accretion disks, 
we present in this study an alternative model for the time  
dependent spectral formation of Comptonized radiation which can reproduce
the basic observational features described above. In \S 2 we describe our model 
and demonstrate the relation between the PSD shapes and the slopes
of the photon energy spectra;  we also emphasize the nature of the
white noise spectra below $\omega_c$ as well as the physics associated
with the low value of this turnover frequency. 
In \S 3 we discuss the issue of the very low-frequency noise and we 
briefly review the relevance of the present model to RXTE observations 
of Cyg X-1 in its low and high states. Finally, we discuss and 
summarize our results in \S 4. 

\section{The Extended Atmosphere Model}

Our study uses as background the model of Chakrabarti \& Titarchuk (1995;
hereafter CT95), i.e. quasi-spherical accretion with a centrifugally
supported shock, which thermalizes the accretion kinetic energy and gives
rise to the hot electrons responsible for the Comptonization of soft
photons. However, we postulate, in addition, that the quasi-spherically 
accreting component behaves like a hot ``atmosphere" of constant electron 
temperature $T_e \sim 50$ keV and, more importantly, with a 
density profile $n(r) \propto 1/r$ in radius, extending to  
$r = r_c\simeq 10^4 ~~R_s$, where $R_s=2GM/c^2$ is Schwarzschild radius. 
This atmosphere is
thought to be the result of preheating of the accretion flow by the
radiation produced at the shock, as discussed by Zeldovich \& Shakura
(1969) and Chang \& Ostriker (1985 and references therein)
(see additional discussion on the issue of this specific density profile in \S 4). 

We assume, for simplicity, that the geometry of this configuration is
spherical and that there is a source of soft photons within the spherical
shock boundary of radius $r_{\rm sh}$. The electron density is
considered to be constant, $n_+$, inside this boundary and $n_+r_{\rm
sh}/4 r$ in the atmosphere outside. The discontinuity of a factor 4 in 
density across the boundary $r_{\rm sh}$ is to account for 
the density jump across the shock. 
The physical size of this cloud is determined by the total
optical depth $\tau_0$.  We have calculated the response of this configuration,
i.e.  the region interior to the shock plus the extended atmosphere, to an
impulsive input of soft photons within the radius of the shock. The
calculations were carried out by a modified Monte Carlo code based on the
method described in Hua \& Titarchuk (1995). Figures 1a and 1b show the
resulting light curves at different energy bands for clouds with 
$\tau_0 = 1, 2$ and 3, $n_+ = 1.6\times 10^{17}$ cm$^{-3}$, 
$r_{\rm sh} = \tau_0 \times 10^{-4}$ light seconds and outer radius of the 
atmosphere $r_c  = 5 \times 10^3 r_{\rm sh} \simeq \tau_0$ 0.5 light
seconds. It is apparent that their shapes have the
form of power laws over the time range $10^{-3}$ s to $\sim$ 1 s, followed
by an exponential cutoff at times of order a few seconds.  For comparison,
the exponential light curves from a cloud with uniform density $n=2 \times
10^{14}$ cm$^{-3}$, radius $r = 1.5 \times 10^{10}$ cm ($\simeq 0.5$
light seconds) the same electron temperature $T_e$ and $\tau_0=2$ 
are also shown (dotted curves). 

The power law form of the light curves, $f(t)$, is the result of 
photons scattering in the extended atmosphere  whose optical path 
has a logarithmic radial dependence
$$
\tau=\int_0^r \sigma_T n_e dr~=~\tau_{\rm sh}[1+0.25\cdot 
\ln (r/r_{\rm sh})] ~~~{\rm for}~~r\geq r_{\rm sh} \eqno(1)
$$  
where the electron density $n_e = n_{+}$ for $r\leq r_{\rm sh}$ and 
$n_{+}r_{\rm sh}/4 r$ for $r\geq r_{\rm sh}$; $\sigma_T$ 
and $\tau_{\rm sh}$ are the Thomson cross section and optical thickness of 
the shock respectively. In the case that the optical 
depth of the extended atmosphere is a few, most photons
escape either unscattered or after one  scattering 
(we exclude in these considerations the photon spread in time over
the short time scales $\sim r_{\rm sh}/c$ associated
with the uniform sphere  at the center of radius $r_{\rm sh}$). 
The light curves of figure 1 present the distribution in the delay times of the 
escaping photons with respect to the light crossing time, 
for photons which have scattered once. The light curve is related 
to the distribution of 
photons over the time of the first scattering $t_1$. The probability 
of scattering in an interval between $t_1$ and $t_1 +dt_1$ is $P(t_1) 
dt_1  = f(\tau) d\tau$ where $f(\tau)$ is the
probability density to scatter between depths $\tau$ and $\tau + d\tau$. In our
case this probability density is constant and equal to $1/\tau_0$, {\it i.e.}
inversely proportional to  the total depth of the atmosphere. Therefore,
$$P(t_1) = {1 \over \tau_0} {d\tau \over dt_1} = {1 \over \tau_0} {d\tau \over dr}
{dr \over dt_1} = {1 \over \tau_0} \sigma n_e(r) c \eqno(2)$$

\noindent For a density distribution which is a power law, {\it i.e.}
$n_e(r) \propto r^{-\alpha}$ (we herein consider the 
case $\alpha =1$) $P(t_1) \propto t_1^{-\alpha}$.
One can establish, by looking at the geometry of  photon trajectories in a  
spherical geometry of radius $R$, that for a photon which has scattered 
at radius $r_1$   the average distance before escape is given by
$$
\langle \ell \rangle = {1 \over 2} r_1 \left[ {R \over r_1} + {1\over 2} 
\left( {R^2 \over r_1^2}-1 \right) \cdot \ln 
\left( {R+ r_1 \over R - r_1} \right) \right]~,
$$
indicating that for small values of $r_1$ the average delay time $t_d$ of photon
escape after one scattering relative to the light crossing time of the 
entire atmosphere of size $R$ is 
$$
\langle t_d \rangle = r_1 \left( 1 - {r_1 \over 3 R} \right) 
\simeq r_1 ~. \eqno(3)
$$
The linearity
between the scattering time $t_1$ and the delay time $t_d$ allows us to substitute
$t_d$ in equation (2) above, leading to a light curve which is a power law in $t$ 
(or better in $t_d$) with index equal to that of the density profile index $\alpha$.

The Monte Carlo calculations shown in figure 1 support the arguments presented above.
The slope of the the light curve for the uniform sphere of small 
scattering depth is close to zero (dotted curve) and  that
of the atmosphere considered here is close to -1; we have also  used a density
distribution appropriate to free-fall  {\it i.e.} $n(r) \propto r^{-3/2}$ with results 
in accordance with the above arguments. It is well known
(Sunyaev \& Titarchuk 1985) that for photons escaping from a source of
finite size, the distribution of number of scatterings is an exponential
i.e. $f(u)\propto \exp(-\beta u)$, where $u$ is the scattering number and
the parameter $\beta$ is inversely proportional to the mean number of
scatterings, $\hat u$, which is greater than $1$ for clouds with total
optical depth $\tau_0 \gax 1 $. Thus, as expected on the basis of this 
argument, the light curve has an exponential turnover at delays longer 
than $1/\beta$ times the scattering time at the largest 
decade of radius. These arguments therefore present the possibility 
that measuring these light curves, e.g. by  computing the resulting PSDs, 
can provide {\it a tool for uncovering the density profile of
the atmosphere} .

As the optical depth of the atmosphere increases, multiple scattering (diffusion)
effects in the extended atmosphere become more important and
the above arguments, which rely on the single scattering approximation, have to be 
reformulated. The power law indices  of the light curves evolve to smaller values;
in fact, in the uniform sphere case and for very large optical depths, as well 
known, the light curves become rising in time (see e.g. Hua \& Titarchuk 1996; 
hereafter HT96). The flattening of the light curves is indicated 
clearly in our our Monte Carlo calculations presented in figures 1a and 1b.
This change in the light curve slopes reflects directly on the resulting PSD spectra,
shown in figures 1c and 1d. It can be seen there that the PSD curves are 
also power laws in Fourier frequency $\omega$, since the Fourier transformation 
of a power-law, $t^{-\alpha}$, is also a power-law in  frequency $\omega^{\alpha-1}$. 
Thus the flattening of the light 
curves leads to steepening of the PSD spectra, which however, as a rule, are 
flatter than those corresponding to the PSD of a  uniform source ($\propto 
\omega^{-2}$; dotted curves), which correspond to exponential-shot light curves. 

Assuming that the temperature of the extended ``atmosphere" is constant, 
a change in its  
total optical depth $\tau_0$, which results in a change of the  
light curve index and a steeper PSD, will manifest itself, in addition, 
in the energy spectrum of the escaping radiation: The increase in optical 
depth will lead to a harder energy spectrum for the emerging photons. 
Therefore, our considerations, supplemented by the Monte Carlo calculations, 
support to the conclusion that,
under the assumption of the existence of an ``atmosphere" with the density 
profile described above,
{\it there should exist well defined correlations between the slopes of the 
escaping photon spectra and those of the PSD}.  We believe that RXTE is 
uniquely suited to search and document the presence of such correlations. 
 
In Figure 2 we plot the energy spectrum (solid curve) resulting from 
the extended atmosphere  with
parameters $kT_e=50$ keV and $\tau_0=3$. It is seen that it is almost
identical to that from a uniform plasma cloud with the same
electron temperature $T_e$ but a
different optical thickness $\tau_0=2$ (dashed curve). For comparison, we
also plot the energy spectrum from a uniform cloud with the same
temperature and $\tau_0=3$ (dotted curve).  It is seen that this spectrum
is harder than the other two. This is because for the same total optical
thickness ($\tau_0=3$), photons in the uniform cloud find it harder to
escape than those in the $1/r$ atmosphere.  This example shows clearly
that spectroscopic analysis alone can not provide complete information
about the source structure and to get a complete picture of the spectral
formation it is necessary to analyze in conjunction the temporal as well as the spectral
data. 

The time information is of course of vital importance for setting the
physical scale of the system. Since it is believed that Comptonization is
the main mechanism for the production of X-rays, measurements of  the time 
lags between two different energies in the X-ray band should give us a 
direct estimate of the density of the region at which Comptonization takes
place. Based on simple arguments concerning the dynamics of accretion
onto compact objects, the associated time lags are expected  be of the order 
of msec for galactic objects. It is worth noting that, for the uniform 
density hot clouds considered almost exlusively todate in the literature,
the time lags between soft and hard photons are constant and independent
of the Fourier frequency $\omega$. Extensive caculations of this type 
({\it i.e.} of scattering in clouds of uniform density) are given
in HT96. For the non-uniform density profile of the extended atmosphere discussed
herein, the shape of the phase lag as a function of the variability
frequency $\omega$ also changes. In Figure 3, we plot the phase 
lags of photons in the 10 -- 20 keV band relative to those in 1 -- 10 keV 
band, for the four cases  depicted in Figure 1, as a function of $\omega$. 
As in Figure 1, the dotted curve corresponds to the configuration of the
uniform cloud  described earlier and its shape is similar to those displayed 
in Figures 14 -- 17 of HT96, except that here the curve extends 
to higher frequencies with
a roughly constant slope instead of a sharp drop. This is because
in present analysis we use
much finer time bins  than before. However, the maximum at
frequency $\sim n_e\sigma_Tc/\tau_0$ (HT96, Eq. 11) is clear and sharp. On
the other hand, the phase lags corresponding to the $1/r$ atmosphere
configuration (solid curves) are much flatter and there is no maximum
present across the entire range of frequencies. It is interesting to point out
that this type of phase lag is hinted in the GINGA observations of Cyg X-1 
(Miyamoto et al. 1988) and also seen in the recent observations of
the high state of Cyg X-1 (Swank et al. 1996). 

As pointed out in HT96, the PSD associated Comptonized photons of 
a given energy, depends strongly
on the energy of the source, soft photons. In Figure 1, the source photons
have a blackbody temperature 2 keV. In Figure 4, we present the PSD curve
(solid curve) in the energy range 10 -- 20 keV resulting from the same
configuration as that of $\tau_0=2$ in Figure 1, but with the source
photons at blackbody temperature $kT_0=0.5$ keV. Compared to the
corresponding PSD curve in Figure 1, it is seen that PSD with lower source
photon energy is steeper, in agreement with the analysis of HT96. 

In Figure 4, we also present one more PSD (dotted) curve, corresponding to
the same light curve as the solid one but for different time ranges and
time bins. For the solid curve the light curve is calculated over a range
of 4 seconds in 4096 bins. For the dotted curve, the light curve is
calculated over 32 seconds in the same number of bins. It is seen that the
PSD turns flat for frequencies below $\omega_c \sim 0.25$ Hz while keeps
parallel to the solid one above $\omega_c$ except at the highest frequency
end, which is obviously due to aliasing (see e.g. Press et al. 1992). 
It is found from examining the light curve that the
time scale corresponding to $1/\omega_c = 4$ seconds is actually the time
scale of the light curve beyond which the latter drops virtually to zero.
Thus we have found a possible physical meaning for the shoulder (break)
frequency $\omega_c$ in the PSD curves, which is common in the PSD of many
sources, namely, it indicates the time scale of the light curve, or 
alternatively the size of the extended $1/r$ atmosphere. The ``white 
noise" below $\omega_c$ reflects the average frequency of the shots while 
the power-law above $\omega_c$ reflects the time structure within one 
single shot. 

\section {The very low frequency noise}

In addition to the fluctuation PSD described above, galactic X-ray sources
exhibit power-law type PSD at frequencies much lower than 
$\omega_c$, so that the white noise component at $\omega < \omega_c$
appears only as a shoulder in the much broader PSD spectrum. We have
thus reasons to believe that this very low frequency component in the PSD
is separate from the one described above. One should bear in mind that at radii
$r \gg r_c$, related to the dynamics associated with this component
in the PSD, the effects of preheating are thought to be small and accretion
proceeds in the form of the standard viscous $\alpha$-disk
(Shakura \& Sunyaev 1973 hereafter SS73). As a result, 
one can consider the fluctuations in the
accretion rate $\dot M = \dot M_0 + \delta \dot m$ to be resulting from  
small random variations in the  viscosity parameter $\alpha$, {\it i.e.} 
$\alpha = \alpha_0 + \delta$. This model was considered by Lyubarskij
(1995), who, by solving the angular momentum diffusion equation,
showed that the resulting accretion rate and hence the luminosity,
has a flicker-type fluctuation spectrum, provided that the characteristic
time of the fluctuations in $\alpha$ are of the order of the viscous
time scales, and the fluctuations in $\alpha$ at different radii
are uncorrelated. 

We provide here a heuristic derivation of this fact for the sake of
completeness of the discussion. The accretion time  $\tau_a$ from 
a radius $r$ in an $\alpha$-disk is

$$
\tau_a=\left[\alpha \left({h\over r}\right)^2\Omega_K\right]^{-1}.\eqno(4)
$$
\noindent
The variation $\delta$ in the viscosity parameter $\alpha$ in an annulus
leads to the mass accretion rate variations $\delta \dot m_i\propto
\delta(r,t)$. In fact the contribution in the mass variation caused by the
viscosity variations is proportional to the annulus area $rdr$, the
surface density $u$ and inversely proportional to the accretion time
$\tau_a$ {\it i.e.}, 
$$ \delta \dot m_i \propto {{\delta\cdot u
rdr}\over{\tau_a}} ~. \eqno(5) $$ 
In different accretion disk regions
(SS73) (region {\bf a}: $p_r\gg p_g$ and $\sigma_T\gg \sigma_{ff}$, region
{\bf b}: $p_r\ll p_g$ and $\sigma_T\gg \sigma_{ff}$, region {\bf c}:
$p_r\ll p_g$ and $\sigma_T\ll \sigma_{ff}$) there are different
dependences of the surface density $u$ and thickness of disk $H$ on
radius. For region {\bf a}, $u\propto r^{1.5}$, $H$ is almost constant and
hence $\tau_a\propto r^{3.5}$ (see Eq. 4) ; for region {\bf b}, $u\propto
r^{-0.6}$, $H\propto r^{21/20}$ and $\tau_a\propto r^{1.4}$; for region
{\bf c}, $u\propto r^{0.75}$, $H\propto r^{9/8}$ and $\tau_a\propto
r^{3.5}$.  Despite the differences in the behavior of the main
parameters over the disk one can check by substituting the above
expressions of $u$, $H$ and $\tau_a$ in Eqs. 4 and 5 that the mass (or
luminosity) variations are the same for all zones {\bf a-c}, namely 
$$
dL~\propto~d\dot
m_i~\propto~{{dr}\over{r}}~\propto{{d\tau_a}\over{\tau_a}}
={{d\omega}\over{\omega}}\eqno(6) 
$$ 
and thus the resulting power spectrum
$p(\omega)$ is 
$$ 
p(\omega)~\propto~\omega^{-1}. \eqno(7) 
$$ 
In other words if the amplitude of the variations in $\alpha$ are the the
same at different radii then the amplitudes of accretion rate
(luminosities) variations are the same at different time scales
(Lyubarskij 1995). 

\section{Discussion and Conclusion}

We have presented above a model which implies corelations between the
spectral and temporal properties of accreting compact objects
and have pointed out that this relation, along with independent
measurements of the electron temperature $T_e$, can allow a complete
specification of the physical parameters of these systems, including the
density profile of the extended atmosphere. 

Much of the present discussion relies on the specific form of the 
density $n(r)$ of the accreting matter as a function of radius $r$. 
Clearly, this is not the free-falling solution customarily used in 
association with spherically symmetric accretion, so a few comments are 
in order. If the presence of the extended atmosphere is due to the effects
of preheating, as suggested in \S 1, then at the edge of this atmosphere, at
$r \simeq r_c$, one would expect the random and rotational 
velocities of matter to be comparable and that the subsequent 
evolution of the accreting fluid in radius to be predominantly governed by 
the removal of its angular momentum. The agent responsible for this process
is considered herein to be the interaction of the fluid with the photons
produced at the base of the extended atmosphere, {\it i.e.} near the 
Schwarzschild radius $R_s$. In order to estimate the effectiveness of this process
we compare the free-fall 
time scale, $t_{ff}$, with that of the viscous time scale, $t_{visc}$, 
assuming that the density has the required profile. Assuming that 
the density profile of the atmosphere  has the form $n(r) = n_0 
(r_0/r)$ between radii $r_0$ and $r_c$, with $r_c \sim 10^3 - 10^4 ~r_0$, 
mass conservation dictates that the the infall velocity, $v(r)$, will also 
have a similar scaling between $r_0$ and $r_c$, {\it i.e.} $v(r) = v_0
(r_0/r)$, yielding for the free-fall time $t_{ff} \simeq r/v(r) = r^2/v_0 r_0$.
The photons can achieve the required angular momentum removal if 
$t_{ff} \simeq t_{visc}$.  

Under the assumed density and velocity scalings, the angular momentum 
of the fluid at a radius $r$ would be ${\cal L} \simeq m_p n(r) v(r) r 
\cdot r^3$. This fluid interacts and transfers its 
angular momentum to photons diffusing from the interior regions of 
the atmosphere which carry substantially smaller angular momentum. 
The specific density profile we have assumed, which implies an  equal 
optical depth per decade of radius, guarantees
that: (a) The angular momentum of the photons, as they traverse each decade 
in radius, is much smaller than the local angular momentum of the fluid. (b) The
photons interact with the fluid at every decade in radius with probability $\sim 1$ 
before their escape, carrying away angular momentum at a rate $\dot 
{\cal  L} \simeq (L/c^2) v(r) r$, where $L$ is the total luminosity of accretion. 
The viscous time scale for the fluid at radius $r$ is then the time required 
for the fluid at radius $r$ to get rid of its angular momentum, {\it i.e.},

$$t_{visc} \simeq {{\cal L} \over \dot {\cal  L}} 
\simeq {n(r) c^2 r^3 m_p \over L} \eqno(8)$$

The luminosity $L$ is given simply by the total accretion rate 
$\dot M$ and the innermost radius of the accretion disk $r_i$, 
{\it i.e.} $L \simeq \beta \dot M c^2 (R_s/r_i) \simeq \beta 
m_p n_0 v_0 r_0^2 c^2 (R_s/r_i) $   
with the parameter $\beta$ allowing for accretion onto the
compact object other than that associated with the spherically accreting
component (most likely associated with a viscous geometrically 
thin disk). Substituting the expression for $L$ onto equation (8) above
yields for the viscous time scale 
$$t_{visc} \simeq {r^2 \over v_0 r_0} {1 \over (R_s/r_i) \beta} =
t_{ff}{1 \over (R_s/r_i) \beta} \eqno(9)$$

\noindent The above expression indicates that the removal of the
angular momentum of the accreting matter by the photons  produced near 
$r \simeq R_s \ll r_c$ can in fact proceed on  dynamical
time scales and thus preserve the assumed density profile provided 
that $(R_s/r_i) \beta \simeq 1$, {\it i.e.} that the total accretion 
rate (including an additional component, most likely from an accretion 
disk) is  $\simeq (r_i/R_s)$ times that associated with the spherically 
symmetric component.
We reiterate that, most likely, this process is possible only for the 
density profile prescribed above ($n(r) \propto 1/r$), 
since this is the only profile which 
allows for significant photon scattering, and hence removal of angular 
momentum, from a large range of radii. 

The presence of the extended atmosphere discussed above, due presumably 
to upstream heating of the accreting matter, may bear relevance 
not only to accreting black holes as discussed herein,  but also to 
accreting neutron stars, with the overall similarity
of their PSD spectra (Van der Klis 1995) not being simply  coincidental. 
It is interesting to speculate that coherent oscillations of 
this configuration (the extented atmosphere) may in fact
be related to the QPO phenomenon, a phenomenon prevalent among all members
of the more general class of accreting compact sources, whether 
neutron stars or black holes. The observed QPO 
frequencies are generally smaller than those
associated with the dynamical motions near a black hole or a neutron star,
and similar to those associated with the outer radius $r_c$ of our model, 
implying the possible relevance of this characteristic time scale 
in these systems too.  
This model may also bear relevance to the recently discovered kHz
QPO's in four LMXBs (Strohmayer \etal 1996, Zhang \etal 1996), as
this is the time scale expected for variability in the centrifugally 
supported shock presumably present at the base of the extended atmosphere, 
especially in objects containing neutron stars rather
than black holes (Titarchuk \& Lapidus 1996). 

The presence of a reflection-by-cold-matter component and fluorescent 
$K_{\alpha}$ line emission in the observed spectra of black hole
candidates and AGNs (e.g. Ebisawa et al 1996) could also be related 
with an extended atmosphere of the type discussed above: A 
significant fraction of
hard photons can in the present case random walk over the entire 
extended atmosphere and the possibility arises that it can 
be reflected by far-away, relatively cold parts of the visous 
accretion disk to produce the required components.

In addition to the above and in accordance with the model of CT95,
one would expect a variation in the temporal properties of accreting 
compact sources while at their different spectral states. As discussed
by CT95, the changes in the spectral states derive from the presence
of a sufficiently large number of soft photons from the viscous accretion
disk to cool the electrons to temperatures at which Comptonization
is ineffective. This manifests as a prominent thermal-like peak 
at energies $\sim 1 $ keV. Under the same conditions, one could expect
the extended hot atmosphere to be absent, a fact which should manifest itself
in the associated PSD by the absence of power at the lowest frequencies.
 
We have presented above a model which implies the existence of
a correlation between the temporal and spectral properties of
accreting compact sources. Our model is 
consistent with certain general characteristics associated with
the PSD of accreting compact sources observed to date, in particular 
with their spectral forms
and the breaks associated with the low frequency turn-overs. 
It is also consistent with the observed frequency dependent 
phase lags between photons of different energies and can also 
account for their photon spectra. This
model is specific enough that we believe can it be meaningfully 
tested by combined spectal - temporal measurements of these sources.
   
The authors acknowledge support by NASA grants NAS-5-32484 and
the RXTE Guest Observing Program. They would also like to thank 
Lorrela Angelini and Jean Swank for informative discussions.
XMH would like to  thank NAS/NRC for support during the course of this study.

\clearpage
\figcaption{
The light curves and PSD for the extended atmosphere of temperature $kT_e=50$ keV
described in the text. Three cases of total Thomson 
optical depths $\tau_0= 1, 2$ and 3 are shown. The electron density has the
form $0.25n_+r_{sh}/r$ for radius $r>r_{sh}$ 
and $n_+$ for $r\leq r_{sh}$, where $n_+=1.6\times 10^{17}$ cm$^{-3}$ and
$r_{sh}=\tau_0 \times 10^{-4}$ light seconds. For comparison, 
the cloud with uniform 
density $n=2 \times 10^{14}$ cm$^{-3}$ with the same temperature and 
$\tau_0=2$ is also shown (dotted curves). The left two panels show the
light curves and PSD for emissions in the energy range $1 -- 10$
keV, while the right two panels for $10 -- 20$ keV.}

\figcaption{
Comparison of the energy spectra of emissions from uniform clouds and
extended atmospheres with the same temperature and optical depth
$\tau_0=3$. The 
solid curves indicate the energy spectrum resulted from same extended
atmospheres as in Figure 1. The dotted curves indicate those from
uniform clouds. The spectrum resulting from an extended atmosphere
of $\tau_0=3$ (solid curve) is almost indistinguishable from that
of a uniform cloud with $\tau_0=2$ (dashed curve). Both spectra have
the same parameter $\beta= 0.334$ although they have different
$\tau_0$.}

\figcaption{
The hard X-ray phase lags resulting from the same atmospheres as in Figure
1. The dotted curve indicates the phase lag resulted from a 
cloud with $\tau_0 =2$ and uniform density $n=2 \times 10^{14}$ cm$^{-3}$.}

\figcaption{
The power spectra resulting from the same shot light curve but different 
time ranges. The emissions are from an extended atmosphere
with $\tau_0=2$ and $kT_e=50$ keV. The density profile is similar to 
the one with $\tau_0=2$ in Figure 1, but with source photons at
blackbody temperature 0.5 keV. The solid curve represents a time range
of 4 seconds while the dotted curve for 32 seconds.}

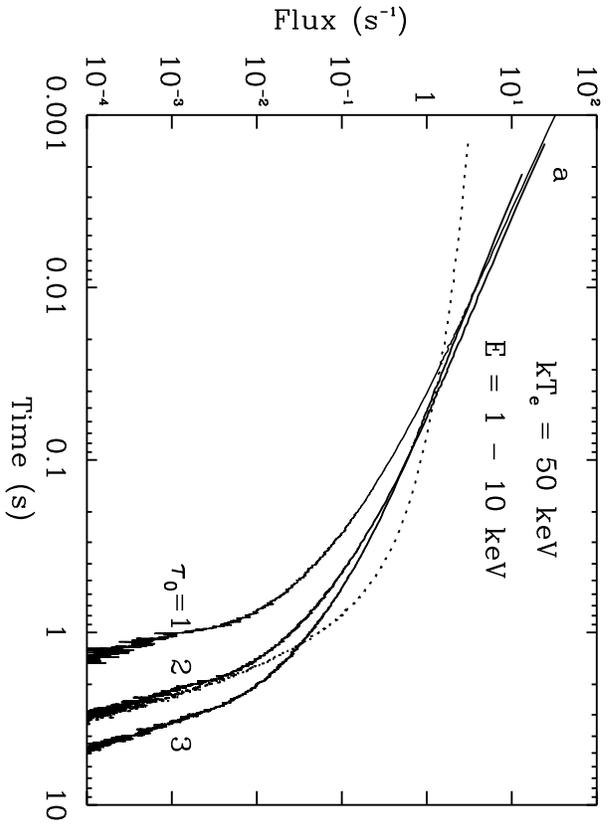
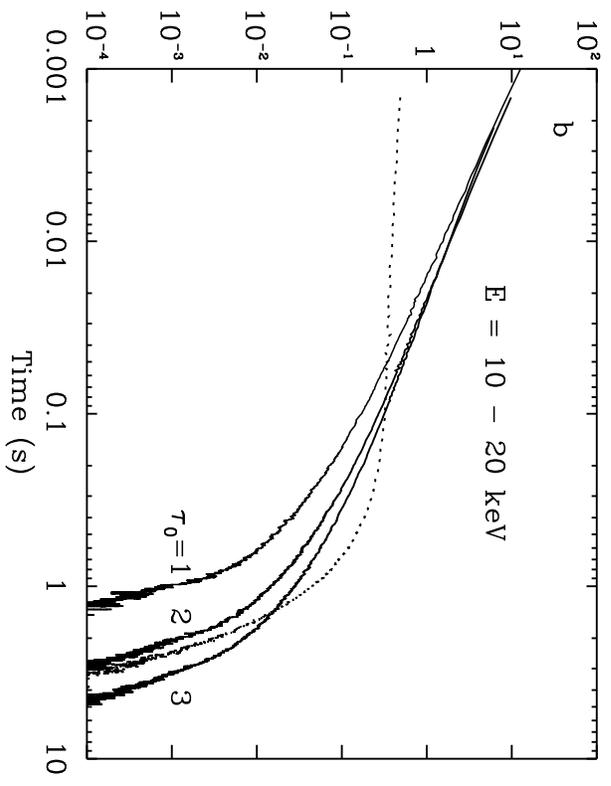
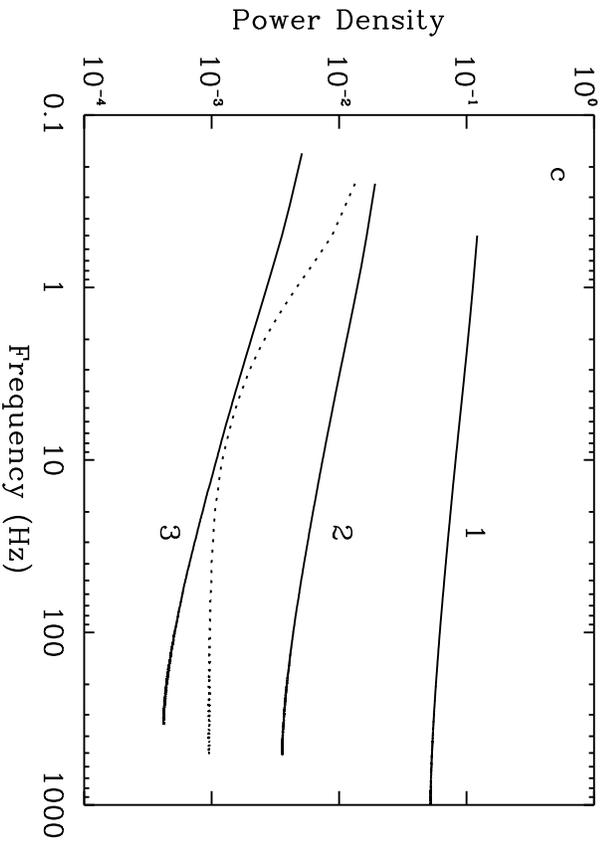
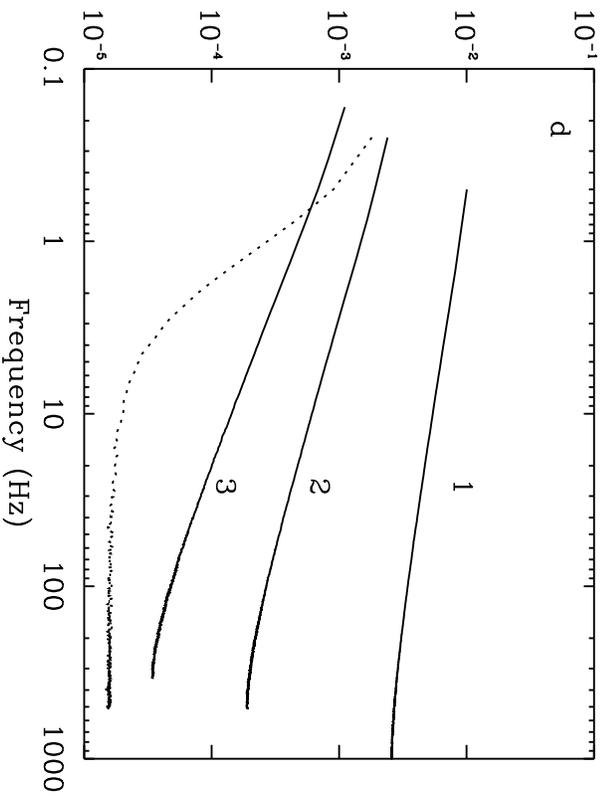
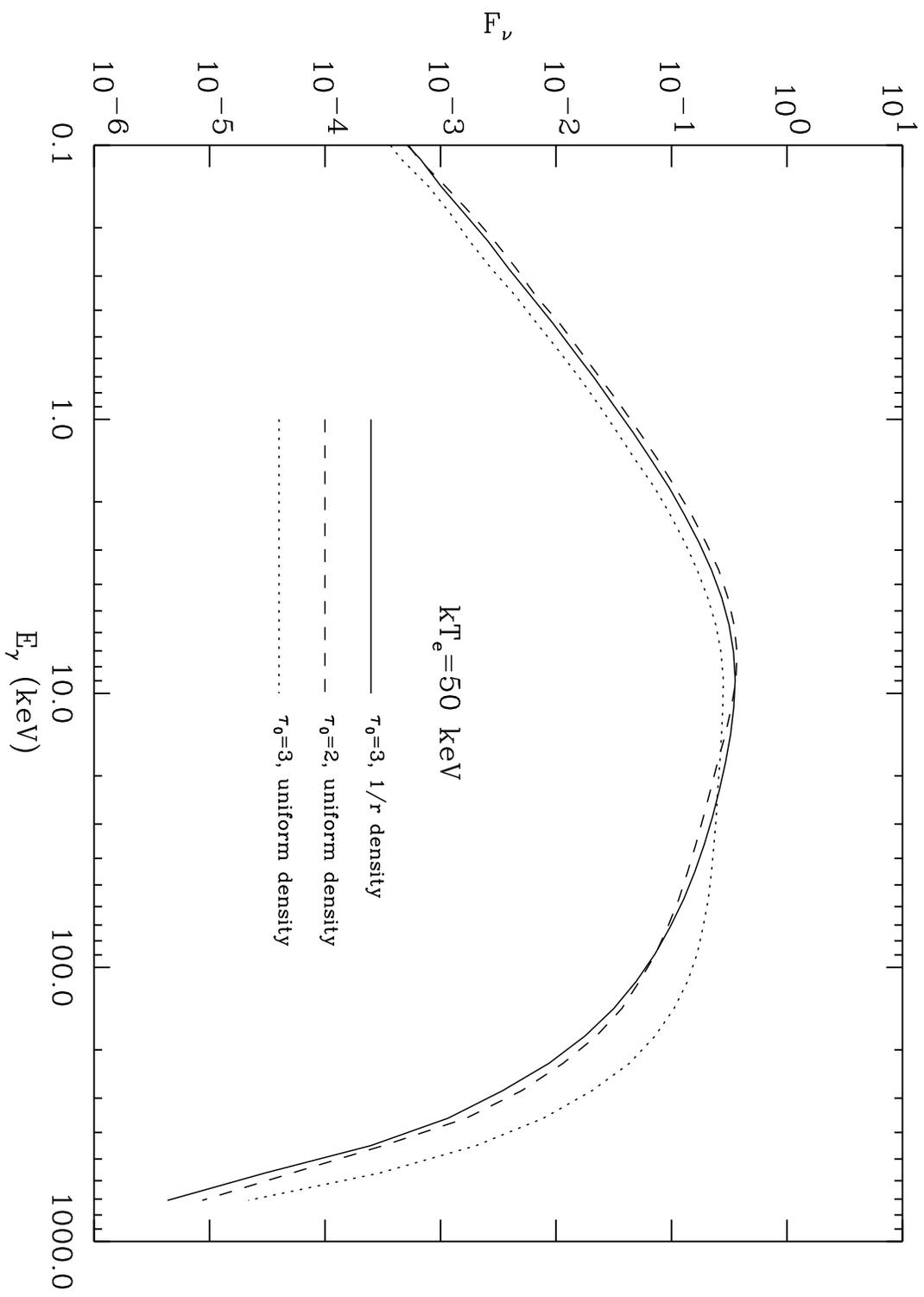
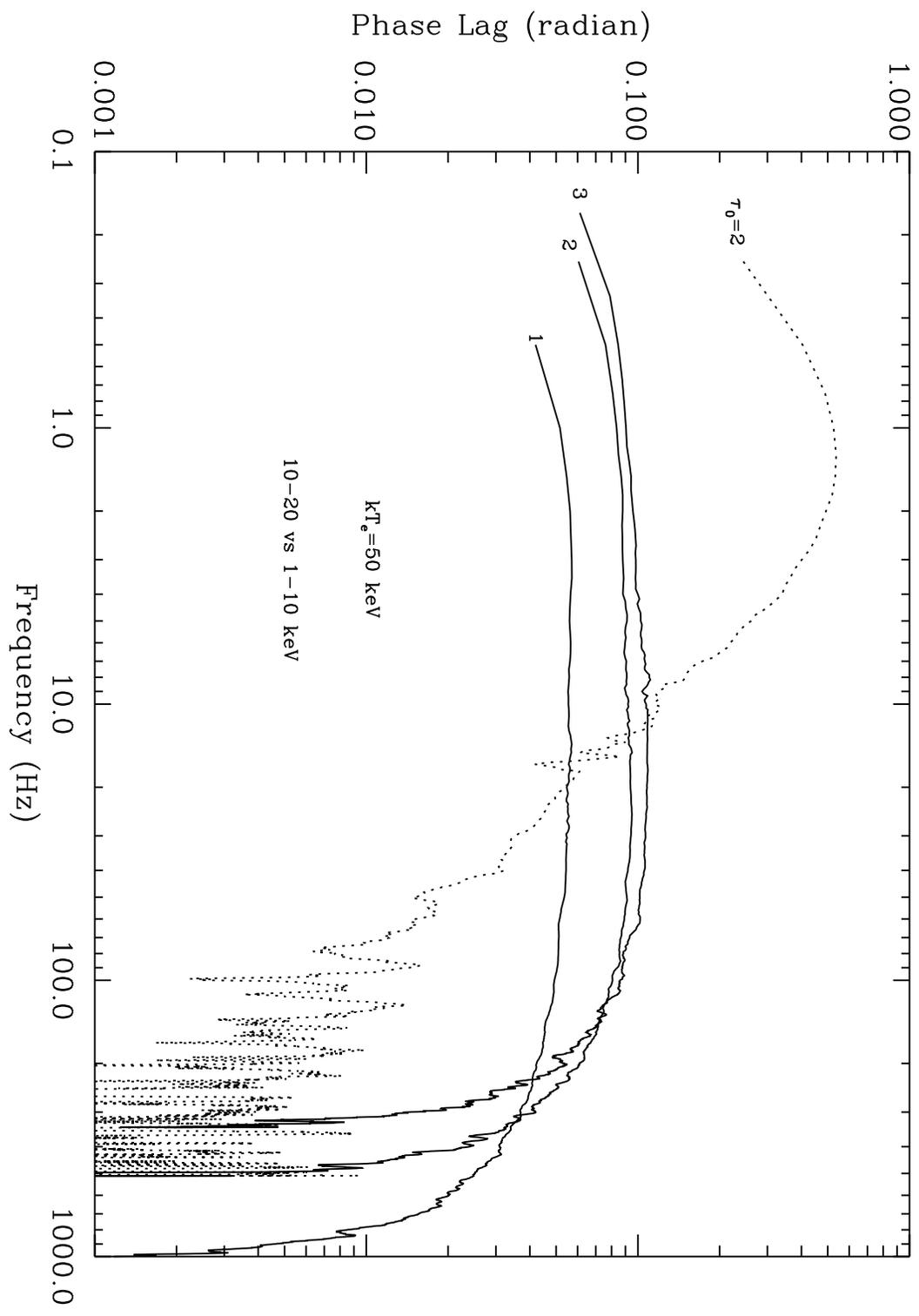
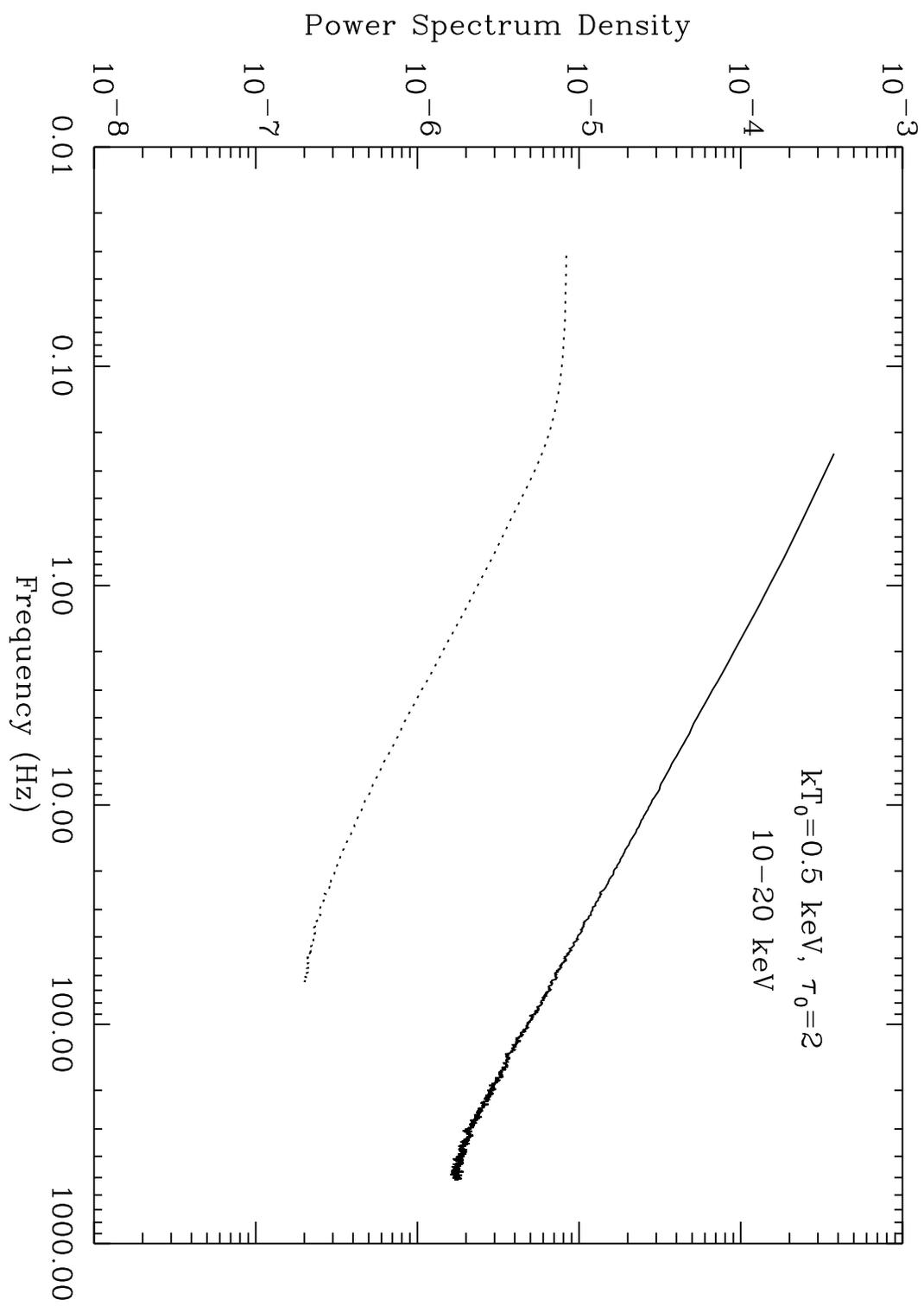
\end{document}